\documentclass[twocolumn,pre,superscriptaddress]{revtex4}
\usepackage{tikz}
\usepackage{graphicx}%
\usepackage{color} 
\usepackage{transparent}
\usepackage{psfrag}
\usepackage{dcolumn}
\usepackage{amsmath,wasysym}
\usepackage{bm} 
\usepackage{amssymb}
\usepackage{latexsym}
\usepackage{hyperref} 
\usepackage{booktabs}
\usepackage{latexsym, bbold}
\begin{document}

\newcommand{\tikzcircle}[2][red,fill=red]{\tikz[baseline=-0.5ex]\draw[#1,radius=#2] (0,0) circle ;}%
\def\bea{\begin{eqnarray}}
\def\eea{\end{eqnarray}}
\def\beq{\begin{equation}}
\def\eeq{\end{equation}}
\def\f{\frac}
\def\k{\kappa}
\def\e{\epsilon}
\def\ve{\varepsilon}
\def\be{\beta}
\def\D{\Delta}
\def\h{\theta}
\def\t{\tau}
\def\a{\alpha}

\def\cDa{{\cal D}[X]}
\def\cD{{\cal D}[x]}
\def\cL{{\cal L}}
\def\cLo{{\cal L}_0}
\def\cLa{{\cal L}_1}

\def\Gv{{\bf G}}
\def\rv{\bf r}
\def\Re{{\rm Re}}
\def\sj{\sum_{j=1}^2}
\def\rk{\rho^{ (k) }}
\def\rek{\rho^{ (1) }}
\def\cek{C^{ (1) }}
\def\rz{\rho^{ (0) }}
\def\rt{\rho^{ (2) }}
\def\rtb{\bar \rho^{ (2) }}
\def\trk{\tilde\rho^{ (k) }}
\def\trek{\tilde\rho^{ (1) }}
\def\trz{\tilde\rho^{ (0) }}
\def\trt{\tilde\rho^{ (2) }}
\def\r{\rho}
\def\tD{\tilde {D}}

\def\s{\sigma}
\def\kb{k_B}
\def\bF{\bar{\cal F}}
\def\F{{\cal F}}
\def\la{\langle}
\def\ra{\rangle}
\def\nn{\nonumber}
\def\up{\uparrow}
\def\dn{\downarrow}
\def\S{\Sigma}
\def\dg{\dagger}
\def\d{\delta}
\def\p{\partial}
\def\l{\lambda}
\def\L{\Lambda}
\def\o{\Omega}
\def\w{\omega}
\def\g{\gamma}

\def\jv{ {\bf j}}
\def\jr{ {\bf j}_r}
\def\jd{ {\bf j}_d}
\def\jdd{ { j}_d}
\def\noi{\noindent}
\def\a{\alpha}
\def\d{\delta}
\def\p{\partial} 
\def\hf{\frac{1}{2}}

\def\la{\langle}
\def\ra{\rangle}
\def\e{\epsilon}
\def\n{\eta}
\def\g{\gamma}
\def\break#1{\pagebreak \vspace*{#1}}
\def\hf{\frac{1}{2}}

\def\rv{{\bf r}}
\def\pc{\phi_c}
\def\rb{\bar\rho}
\title{Two step melting of the Weeks- Chandler- Anderson system in two dimensions}

\author{Shubhendu Shekhar Khali} 
\email{shubhendushekhar@iisermohali.ac.in}
\affiliation{Department of Physical Science, Indian Institute Of Science Education and Research Mohali,  Punjab, India 140306}

\author{Dipanjan Chakraborty}
\email{chakraborty@iisermohali.ac.in}
\affiliation{Department of Physical Science, Indian Institute Of Science Education and Research Mohali, Punjab, India 140306}

\author{Debasish Chaudhuri}
\email{debc@iopb.res.in}
\affiliation{Institute of Physics, Sachivalaya Marg, Bhubaneswar 751005, India}
\affiliation{Homi Bhaba National Institute, Anushaktigar, Mumbai 400094, India}

\date{\today}

\begin{abstract}
We present a detailed numerical simulation study of a two dimensional system of particles
interacting via the Weeks- Chandler- Anderson potential, the repulsive part of the Lennard- Jones potential. With reduction of density, the system shows a two- step melting: a continuous melting from solid to hexatic phase, followed by a a first order melting of hexatic to liquid. The solid- hexatic  melting is consistent with the Kosterlitz-Thouless-Halperin-Nelson-Young (KTHNY) scenario and shows dislocation unbinding. The first order melting of hexatic to fluid phase, on the other hand, is dominated by formation of string of defects at the hexatic- fluid interfaces.
\end{abstract}

\maketitle


\section{Introduction}
\label{sec:intro}
The high density phase of mono-dispersed particles in two dimensions
(2d) can not support long ranged positional order due to thermal
fluctuations~\cite{Mermin1966, Mermin1968, Halperin2019}. However, the
pair correlation function in it does not decay exponentially, unlike
in the fluid phase.  It shows a power- law decay, characterizing a
quasi- long- ranged positional order.  In addition, the correlation
between the orientation of the bonds between neighbours, the so-called
bond- angle correlation, turns out to be long- ranged. It remains
finite at the longest separations in the system. The quasi- long-
ranged translational order, and the long- ranged bond- orientation
order characterise the 2d- solid.

On the other hand, a system of particles interacting via a short- ranged repulsion, e.g.,
the sterically stabilized colloids, gets into a homogeneous and
isotropic fluid phase at low densities. This phase is devoid of both
the translational and orientational order discussed above.  Apart from
these two phases, an intervening third phase, the hexatic phase, is
possible in 2d systems.  This phase is characterized by a short
ranged positional order and a quasi- long ranged bond- orientational order.

The melting of 2d solid attracted an enormous amount of attention in
literature~\cite{Grunberg2014,Glaser1992,Alder:1962, Lee:1992,
  Zollweg:1992,Kosterlitz1973,Halperin1978,Nelson1979,Young1979}.  The
Kosterlitz-Thouless-Halperin-Nelson-Young (KTHNY)
theory\cite{Kosterlitz1973,Halperin1978,Nelson1979,Young1979}
predicted a two stage continuous melting transition in 2d, from a 
quasi- long ranged ordered (QLRO) solid
to fluid via an intervening hexatic phase. Within this scenario both
the melting are mediated by defect unbinding -- dislocation unbinding
for solid- hexatic melting, and disclination unbinding for hexatic-
fluid melting.  In contrast, early simulations showed signatures of a
first order melting of 2d solid~\cite{Alder:1962, Lee:1992,
  Zollweg:1992}.  The mean field theories also predicted a first order
transition~\cite{Ramakrishnan1979, Denton1989, Zeng1990, Ryzhov1995}.
A later and more careful constrained Monte-Carlo simulation of hard
disk particles showed presence of signatures of both a continuous melting
transition, and a first order transition~\cite{Sengupta2000}.
This simulation did not allow the formation of defects but counted all
the Monte-Carlo moves that potentially formed them, to calculate the
defect- core- energy, and the unrenormalized Young's modulus. These
two quantities along with the KTHNY recursion relation predicted the
precise melting point of the 2d hard- disk solid at
$\eta = \f{\pi}{4} \r d^2 = 0.719$~\cite{Sengupta2000, Nielaba2004}, where $\r=N/A$
is the particle density and $d$ denotes the hard- disk diameter.  
However, the direct numerical simulations of hard disks also showed a clear signature of
phase coexistence at a lower $\eta$~\cite{Jaster:1999, Bernard:2011}. 
For finite systems, the free energy is not necessarily convex unlike in the thermodynamic limit,  
and the equation of state may form a stable loop due to interfacial free energies.
Such a loop was observed in the $P-\eta$ diagram, 
with a coexistence interval at a density lower than the solid melting point,
characterizing a first order transition~\cite{Jaster:1999,Sengupta2000,Bernard:2011}.  
It is only recently that for hard disks the presence of a continuous melting of solid, 
and the first order transition have been reconciled~\cite{Bernard:2011, Engel2013}.  
Ref.\citenum{Bernard:2011} showed a two- step melting in hard disks, 
with a continuous (KTHNY) solid- hexatic melting at 
$\eta \approx 0.72$, and a  first order hexatic- fluid melting with phase-
coexistence in the range of packing fractions
$0.700 \lesssim \eta \lesssim 0.716$.

In hard disks elastic and entropic effects have the same origin. For
more general potentials, however, they are not strictly related, and other
scenarios of phase transition may emerge~\cite{Bernard:2011,
  Hajibabaei:2019, Kapfer:2015ca}.  Using $r^{-n}$ interactions,
Ref.\citenum{Kapfer:2015ca} showed how the boundary of solid to
hexatic continuous melting, and the fluid- hexatic coexistence region
depends on the changing values of the steepness of the potential, $n$.  At large $n$ the two- step
melting, like in hard- disks, survives. At a smaller $n$, defects
become ubiquitous and the nature of hexatic changes. For
$n \lesssim 6$, the fluid- hexatic transition becomes continuous.

The numerical implementation of a potential energy of the form
$1/r^n$, involves a cutoff distance $r_c$ beyond which particles do
not interact.  In general, a discontinuity in the interaction force is
encountered around the cutoff. This could be made small by increasing
the cutoff range. However, a change in $r_c$ changes the amount of collisions,
modifying the value of the pressure~\cite{Kapfer:2015ca}.  Such an
issue does not arise for the Weeks- Chandler- Anderson (WCA)
potential, the repulsive part of the Lennard-Jones potential, for
which the cutoff $r_c$ is chosen at the potential minimum, where the
interaction force vanishes~\cite{Weeks:1971,
  Chandler1983}. Surprisingly, despite its importance in the modeling
of soft matter systems including polymers, colloids and
fluids,~\cite{Weeks:1971, Chandler1983, Grest1986b, Frenkel2002,
  Kroger2004} studies of phase transitions in WCA- system, even in
three dimensions, have found only limited attention~\cite{Ahmed:2009,
  DeKuijper1990, Hess1998}.  A few early attempts with 2d phase
transition of the WCA- particles~\cite{Toxvaerd1983a, Glaser1990,
  Glaser1992} showed a loop in the pressure- density curve, and also  
formation of topological defects. However, as far as we know, the detailed nature of the
melting of the WCA- solid remains to be fully understood.

In this paper, we consider phase- transitions
in the 2d system of particles interacting via the WCA- potential.
Performing large scale molecular dynamics simulations, and careful
analysis, we investigate the melting transitions in this system with
decreasing density.  We find a continuous solid- hexatic melting,
followed by a first order melting of the hexatic to the fluid
phase. The solid melting is associated with dislocation unbinding and
 shows signatures of KTHNY transition. The first order hexatic melting is associated with a Maxwell's
loop in the equation of state, and a clear hexatic- fluid phase coexistence.

The rest of the paper is organized as follows. In
Sec.~\ref{sec:model_simulation} we describe the model and
the numerical simulations. The identification of  different phases
of the system,  the two melting transitions, and formation of topological
defects are described in Sec.~\ref{sec:results_and_discussion}.  
Finally we summarize our main results and conclude in Sec.~\ref{sec_discussion}.

\begin{figure*}[!ht]
  \centering
  \includegraphics[width=\linewidth]{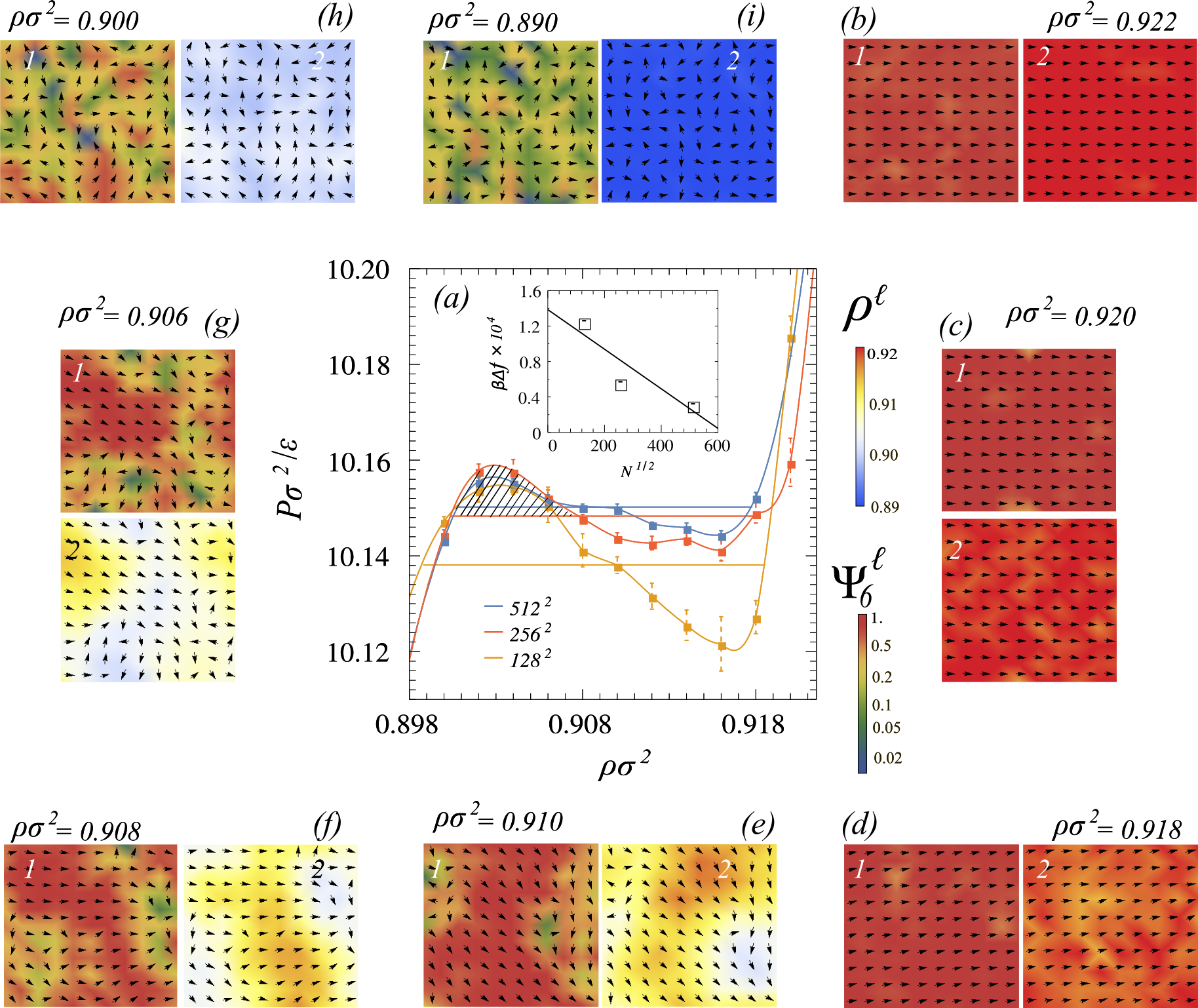}
  \caption{($a$)~Equation of state: Plot of the thermodynamic pressure as a
    function of density for three different system sizes $N=128^2,\, 256^2,\, 512^2$ indicated
    in the legend. The loops in the equation of state are clearly visible. 
    The horizontal lines indicate the Maxwell's equal area construction. 
    The coexisting densities for $N=128^2$ are $\r_1 \s^2=0.899$ and $\r_2 \s^2=0.919$, 
    for $N=256^2$ are
    $\r_1 \s^2=0.900$ and $\r_2 \s^2=0.918$ and for $N=512^2$ are $\r_1 \s^2=0.901$
    and $\r_2 \s^2=0.918$. The shaded area in the $N=256^2$ plot corresponds to the interfacial free
    energy $\beta \D f$ between the majority and the minority
    phases. The inset depicts the scaling of the interface free energy
    with particle number $\beta \D f \sim N^{-1/2}$.
    ($b$)--($i$)~Plots of local hexatic order and density using the $N=256^2$ system at $\r \s^2$ values indicated above each plot. 
    They show heat- maps of the magnitude of the coarse-grained hexatic order 
    $\psi_6^\ell(\bf{r})$ for a single configuration
    (plots labelled with numeral $1$) and the scalar density field
    $\r^\ell(\bf{r})$ averaged over $200$ configurations ( plots
    labelled with numeral $2$) for the whole system. The coarse graining is performed by averaging over  sub-systems of size $\ell^2 = L_x/10 \times L_y/10$.
    Superimposed on
    these fields are shown the orientations of $\psi_6^\ell$ denoted by arrows. 
    }
  \label{fig:qualitative_description}
\end{figure*}

\section{Model and simulation}
\label{sec:model_simulation}
We consider a two-dimensional system of 
$N$ particles interacting via the WCA potential
$U(r_{ij})=4\epsilon[(r_{ij}/\sigma)^{-12}-(r_{ij}/\sigma)^{-6}] +
\epsilon$
for separation $r_{ij}<r_c= 2^{1/6}\sigma$, and $U(r_{ij})=0$
otherwise~\cite{Weeks:1971}.  Here $r_{ij}$ denotes the separation
between $i$-th and $j$-th particle.  The choice of cutoff separation
$r_{c} $ is made such that the repulsion between particles vanishes at
that separation.  The energy, length and time scales are set by $\e$,
$\s$, and $\t = \s \sqrt{m/\e}$, respectively. The mass of the
particles is chosen to be $m=1$.  We use $N=65536$ particles in a
simulation box of size $A=L_x \times L_y$.  At a mean density
$\r = N/A$, a triangular lattice configuration has a lattice parameter
$a$ obeying $a^2=2/\sqrt{3} \r$.  
The separation between the
consecutive lattice planes is given by $a_y=\sqrt{3} a/2$. In our
simulations, we use $L_x = \sqrt{N} a$ and $L_y = {\sqrt N} a_y$, and
the periodic boundary condition.  
The density is controlled by changing the box size. 
We perform molecular dynamics simulations using the standard Leap-Frog
algorithm~\cite{Frenkel2002} with step-size $\d t =0.001\, \t$, in the
presence of a Langevin heat bath characterized by an isotropic
friction $\g = 1/\t$ fixing the temperature $T = 1.0\,\e/\kb$.  At high
densities, the initial configuration is chosen to be a triangular
lattice.  The system is equilibrated over $10^8$ steps, following which
statistics is collected over a further $10^7$ steps.  All simulations
were performed using massively parallel home-grown codes implemented
on Graphics Processing Units (GPU).

\section{Results and Discussion}
\label{sec:results_and_discussion}
In this section we study the phase transition using the equation of state, density fluctuations,
and the hexatic and solid order parameters.

\subsection{Equation of state}
The first evidence of a first order transition comes from the
pressure-density diagram of the system shown in
Fig.~\ref{fig:qualitative_description}. The thermodynamic pressure is
determined from the molecular dynamics trajectories using the virial
expression
\begin{equation}
 P = \kb T \rho + \frac{\rho}{2 N} \sum_i^N \sum_{j>i}^N \bf f(r_{ij}) \cdot \bf r_{ij} 
\label{eq:press}
\end{equation}
where $\bf f(r_{ij}) $ is the interaction force and $\bf r_{ij}$ is
the inter-particle separation. At the fixed temperature $\kb T=1$,
the variation of pressure with density is shown in
Fig.~\ref{fig:qualitative_description}~($a$) for three system sizes:
$N=128^2$, $256^2$ and $512^2$. 
They clearly show Mayer- Wood loops in the equation of states~\cite{Mayer1965}.
While interfacial free energy associated with phase coexistence in 
first order transition explains the loop~\cite{Mayer1965}, the converse is not always true~\cite{Alonso1999}. 
The interfacial free energy at phase- coexistence is expected to 
scale as the interface- length, $\D F \sim N^{1/2}$, in two dimensions. On the other hand, if present, $\D F$ in continuous transition gets independent of $N$ for large system size, and the equation of state becomes monotonic~\cite{Alonso1999, Lee1991}. 
A Maxwell construction on the equation of state suppresses the interfacial effect, and gives the boundary densities $\r_1 \s^2 = 0.900$ and $\r_2 \s^2 =0.918$ for the coexistence interval in system size $N=256^2$. The interval is only weakly dependent on the system size~(Fig.~\ref{fig:qualitative_description}~($a$)\,).
Integrating the equation of state over the shaded region in Fig.~\ref{fig:qualitative_description}~($a$)
we obtain the interfacial free energy per particle  $\D f = \D F/N = \int_{\r_1}^{\r_2} d\r\, (P/\r^2)$.  It shows a scaling form $\D f \sim 1/\sqrt{N}$ as in the first order phase transition~(see  the inset of Fig.~\ref{fig:qualitative_description}~($a$)\,). 

We further follow the phase transition utilizing the 
coarse- grained vector field of the hexatic bond orientational order
$\psi_6^\ell(\bf{r})$, and the scalar field of the local number density 
$\r^\ell(\bf{r})$. 
Unless specified otherwise, here and in the rest of the paper we present results for a $N=256^2$ system.
The fields are obtained by averaging over sub-systems of size
$\ell^2=\ell_x\times \ell_y$ with $\ell_x=L_x/10$ and $\ell_y=L_y/10$.
In obtaining the coarse- grained $\psi_6^\ell(\bf{r})$, we use the hexatic order of each particle 
$\psi^k_6=(1/n)\sum_{j=1}^n e^{i\, 6 \theta_{kj}}$, where  
the angle $\theta_{kj}$ denotes the orientation of the bond vector ${\bf r}_{kj}$ between the
test particle $k$ and its topological neighbor $j$ with respect to the $x$-axis, 
and $n$ is the number of topological neighbors. 

In Fig.~\ref{fig:qualitative_description}~($b$)--($i$) we show the magnitude 
of the coarse-grained field $\psi_6^\ell(\bf{r})$ for
a single configuration, and the corresponding time-averaged scalar
density field $\r^\ell(\bf{r})$ using heat maps.  
The orientations of the $\psi_6^\ell(\bf{r})$ vector are denoted by arrows
in these figures. The configurations in the coexistence interval, as pointed out 
by the Maxwell construction, are shown in Fig.~\ref{fig:qualitative_description}~($d$)--($h$).

For densities above $\r \s^2=0.920$
(Fig.~\ref{fig:qualitative_description}~($b$) and ($c$)), the magnitude of
$\psi_6^\ell(\bf{r})$ remains uniform throughout the system and its
orientations remain aligned
along the $x$-axis,  suggesting a long- ranged hexatic order. 
The time-averaged density field shows little fluctuation. 
The mean density $\r \s^2=0.918$ starts to 
show appearance of low- density and low hexatic order droplets in the otherwise ordered background of large $\psi_6^\ell(\bf{r})$, and $\r^\ell(\bf{r})$~(Fig.~\ref{fig:qualitative_description}~($d$)). 
As the density is decreased further, the system- wide orientational order of
hexatic field $\psi_6^\ell(\bf{r})$ starts to dwindle. 
In Fig.~\ref{fig:qualitative_description}~($e$) and ($f$), still a system spanning band of high hexatic order maintaining a high degree of orientational correlation is observed, coexisting with surrounding low- density fluid domains characterized by low hexatic order and randomized orientations. The  associated density field further highlights the coexistence of high and low density regions,
 corresponding to the high and low hexatic order, respectively. The curved interfaces between the hexatic and fluid regions can be clearly seen from these plots. 
At even smaller density, $\r \s^2 = 0.906$ in Fig.~\ref{fig:qualitative_description}~($g$),
the largest hexatic cluster can not span the system any more. Regions of significantly low hexatic order $| \psi_6^\ell({\bf r}) | \lesssim 0.05$, which already started to appear at $\r \s^2 = 0.910$,  proliferates further at lower densities.
At $\r \s^2 = 0.900$ in Fig.~\ref{fig:qualitative_description}~($e$), we observe small hexatic clusters coexisting with disoriented fluid regions having low hexatic order. The local density plot captures the density fluctuations. Finally, at the smallest density of $\r \s^2 = 0.890$, the whole system shows a loss of hexatic order, and $\r^\ell(\bf{r})$ displays a uniform profile corresponding to the  homogeneous fluid.

\subsection{Solid melting}
\label{sec:solid_melting}
Having established a coexistence interval of $0.900 \leq \r \s^2 \leq 0.918$,
here we proceed to investigate the solid melting using the 
structure factor, solid- order parameter, the positional order and its 
correlation function. 

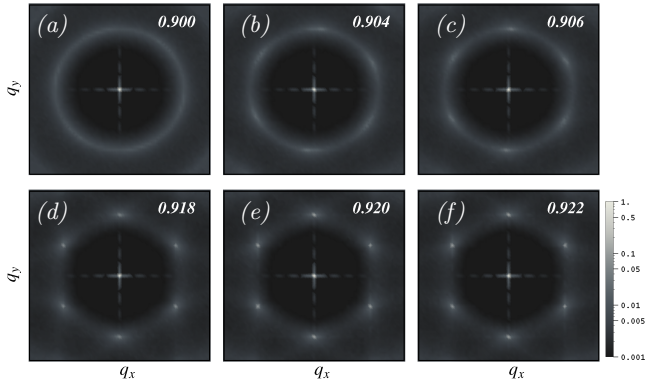
\begin{figure}[!t]
\centering
\begin{tikzpicture}
\node[above,right] (img) at (0,0)
{\includegraphics[width=\linewidth]{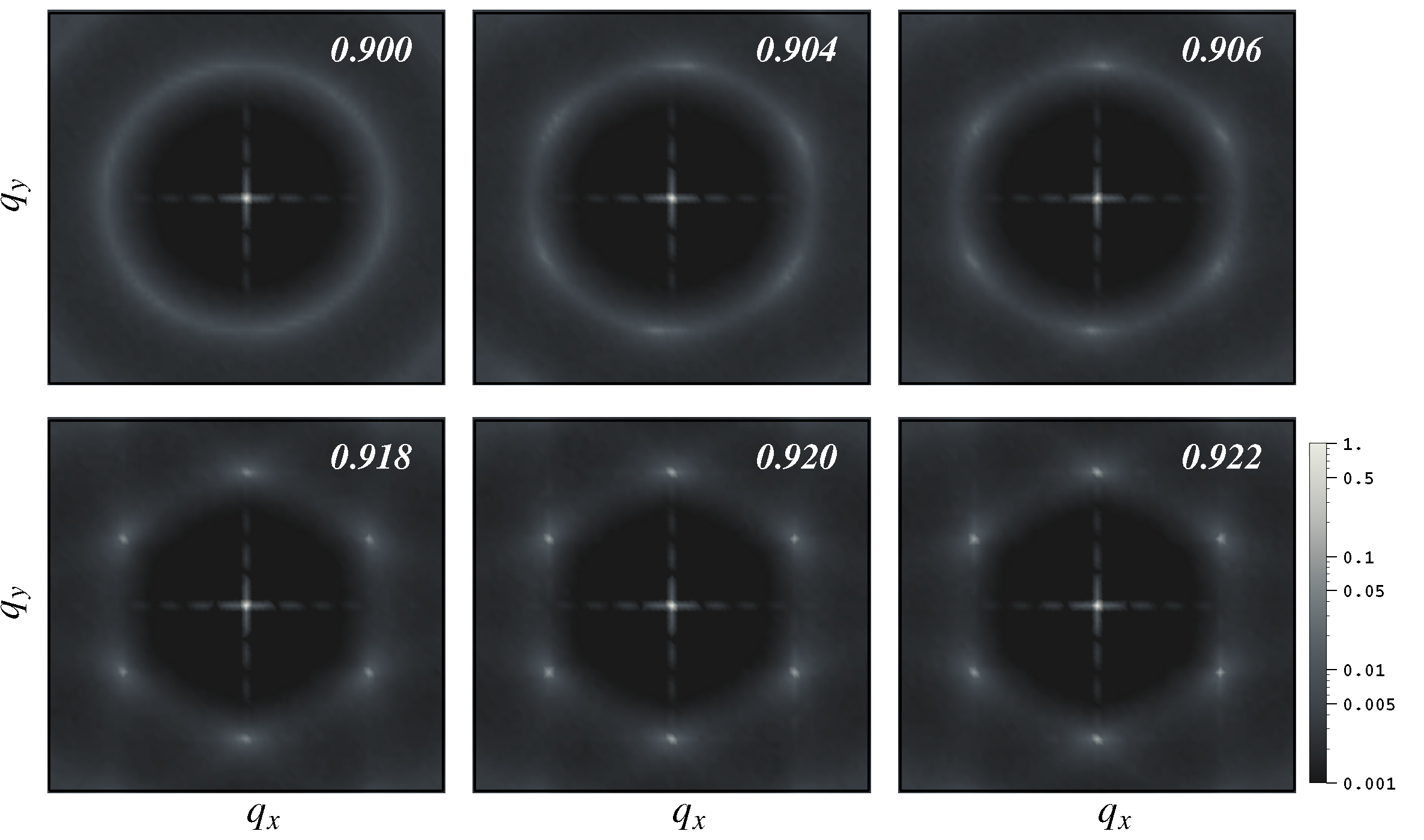}};
\node[text=white] at (0.7,2.2) {{\it  (a)}};
\node[text=white] at (3.4,2.2) {{\it  (b)}};
\node[text=white] at (6.,2.2) {{\it  (c)}};
\node[text=white] at (0.7,-0.3) {{\it  (d)}};
\node[text=white] at (3.4,-0.3) {{\it  (e)}};
\node[text=white] at (6.,-0.3) {{\it  (f)}};
\end{tikzpicture}
\caption{The static structure factor for different densities of the
  system as indicated in the legend.  $\la \psi_{\bf q} \ra$ is
  calculated in the Fourier plane with resolution interval of
  $0.01\s^{-1}$. With reduction of density, the plots show characteristic features of the solid, hexatic, 
  and isotropic fluid phase. }
  \label{fig:skxky}
\end{figure}

\subsubsection{Structure factor}
The static structure factor, defined as
$\la \psi_{\bf q}\ra=N^{-1} \la \r_{\bf q} \r_{-\bf{q}}\ra$ where
$\r_{\bf q}=\sum_{j=1}^N e^{i \bf{q}\cdot \bf{r}_j}$ with
$\r^{*}_{\bf q}=\r_{-\bf q}$ is shown in
Fig.~\ref{fig:skxky} at different densities. It clearly
distinguishes between the solid, the hexatic and the fluid phase.  In
the solid phase, the six quasi- Bragg peaks in $\la \psi_{\bf q}\ra$
capture the characteristic six- fold symmetry corresponding to the
underlying triangular lattice. 
In contrast, in the fluid phase one obtains the characteristic
ring structure of $\la \psi_{\bf q}\ra$ capturing the isotropy of the system~(Fig.~\ref{fig:skxky}\,($a$)).
At the intermediate densities, the intensity modulation on the
fluid-like ring in $\la \psi_{\bf q}\ra$ shows a six- fold symmetry
but with broadened peaks~(Fig.~\ref{fig:skxky}\,($b$), ($c$)).  This is a
characteristic of the hexatic phase~\cite{Chaikin2012}.

\begin{figure}[!t]
\centering
\includegraphics[width=8cm]{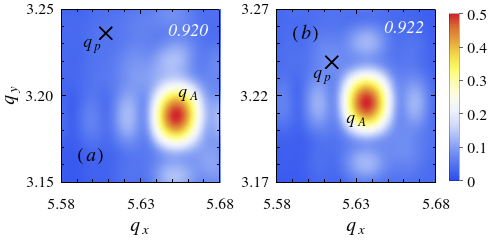}
\caption{Plots of $\la \psi_{\bf q}\ra$ around a quasi- Bragg peak at densities $\r \s^2=0.920$~($a$) and $0.922$~($b$). Its actual location ${\bf q}_A$, the position of the largest value of $\la \psi_{\bf q}\ra$~(red patch), is  shifted from the expected peak- position ${\bf q}_p$ for a perfect triangular lattice~(cross).  The shift is larger at lower density. 
  }
  \label{fig_shift}
\end{figure}

 Here, it is important to note that the actual position of peaks ${\bf q}_A$ in the structure factor $\la \psi_{\bf q}\ra$ of solid are shifted from the expected peak- positions of a perfect triangular lattice, ${\bf q_p}=(0,\pm 2\pi/a_y(\r))$,  $(\pm 2 \pi/a(\r),\pm \pi/a_y(\r))$.  
 This is shown in Fig.~\ref{fig_shift}, focussing on a single quasi- Bragg peak. 
 The shift is both in amplitude and orientation, and the amount of shift depends on the system density.  Similar shifts were previously observed in other models~\cite{Bernard:2011, Li2019}. As it has been pointed out before~\cite{Li2019}, it is important to use the actual positions of the quasi- Bragg peaks ${\bf q}_A$, instead of $\bf{q_p}$, in all calculations involving positional order and correlation.

\begin{figure}[!ht]
\centering
\includegraphics[width=7cm]{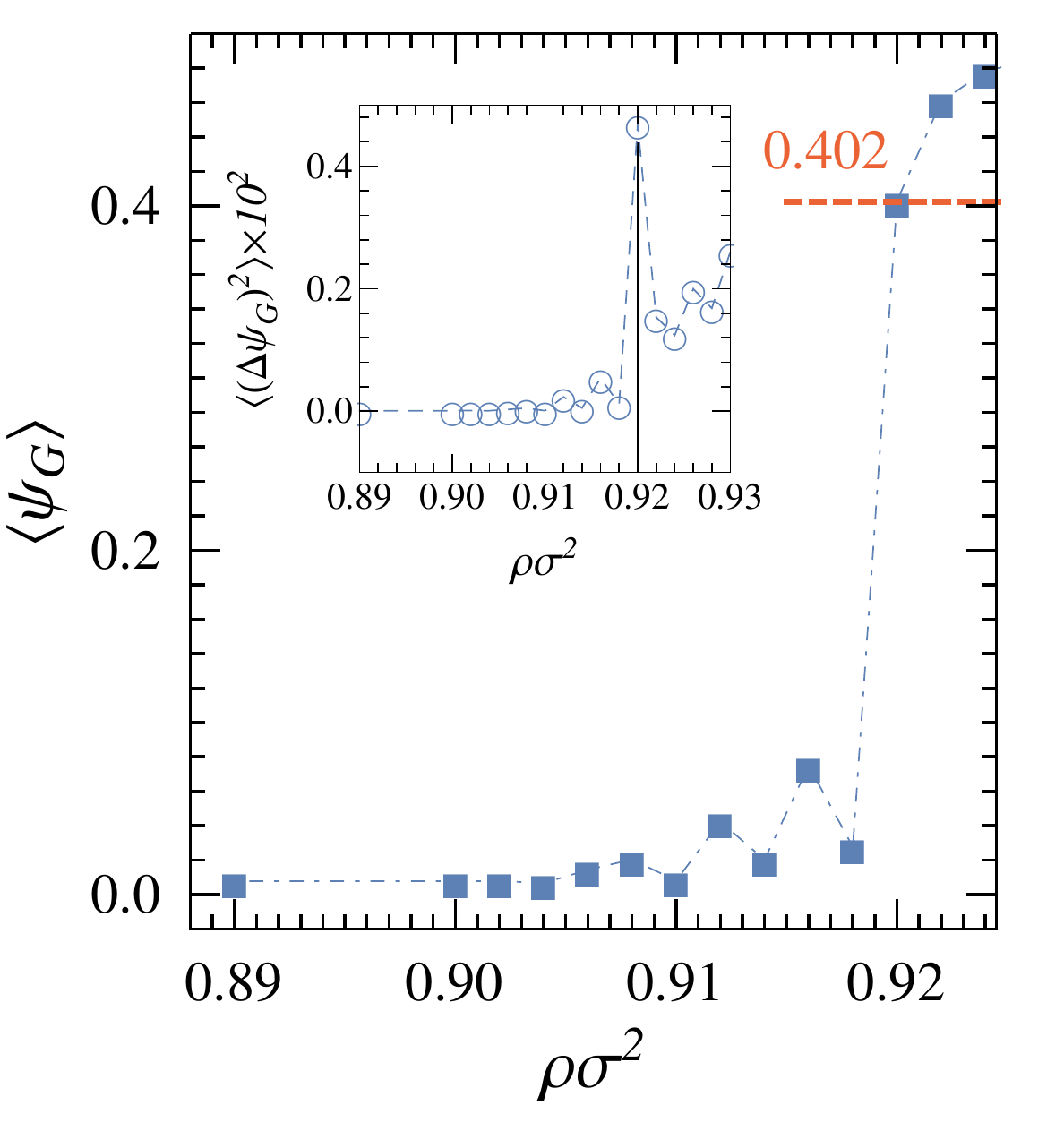}
\caption{Plot of the solid order parameter ($a$) and its fluctuation (inset)
  as a function of density. The peak of the order parameter fluctuation
  appears at a density $\r \s^2=0.920$. 
  }
  \label{fig:solid_op}
\end{figure}

To quantify the solid melting transition, we use the solid order
parameter $\langle \psi_{\Gv}\rangle$, which is an average of $\la \psi_{\bf q}\ra$ over 
the  six quasi- Bragg peaks at $\Gv:=\{{\bf q}_A\}$. 
In Fig.~\ref{fig:solid_op} we show the variation of $\la \psi_{\Gv} \ra$ 
as a function of the mean density
of the system. At very high densities, $\langle \psi_{\Gv}\rangle$
remains large, and drops sharply near $\r \s^2 \approx 0.92$ to 
vanishingly small values. The fluctuations of the order parameter quantified by the
mean squared deviation $\la \D \psi_{\Gv}^2 \ra$ shows
a pronounced maximum at $\r \s^2 = 0.920$\,(see the inset of
Fig.~\ref{fig:solid_op}\,), identifying the melting point of the solid.
Here we emphasize that the melting point $\r \s^2 = 0.920$ remains above the
interval of phase coexistence identified in Fig.~\ref{fig:qualitative_description}~($a$).   

\begin{figure}[!ht]
\centering
\includegraphics[width=7cm]{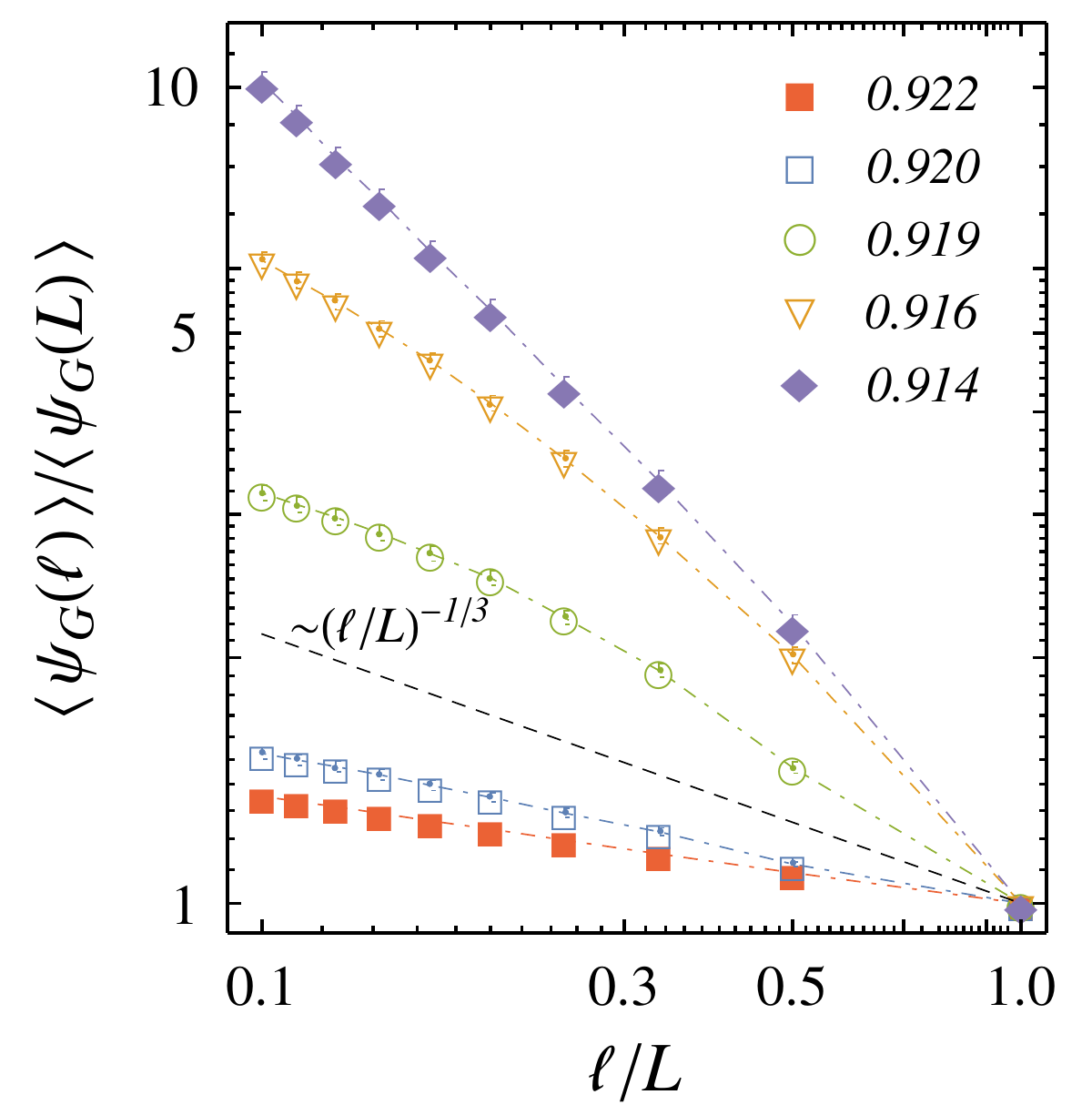}
\caption{ Finite size scaling of solid order parameter $\la \psi_{\Gv}(\ell) \ra$ shown in
  log- log scale. Here $\ell$ denotes block-sizes. 
  The dashed black line is the plot
  of $(\ell/L)^{-1/3}$.  }
  \label{fig_psiG2scaling}
\end{figure}

\begin{figure}[!ht]
    \centering
    \includegraphics[width=7cm]{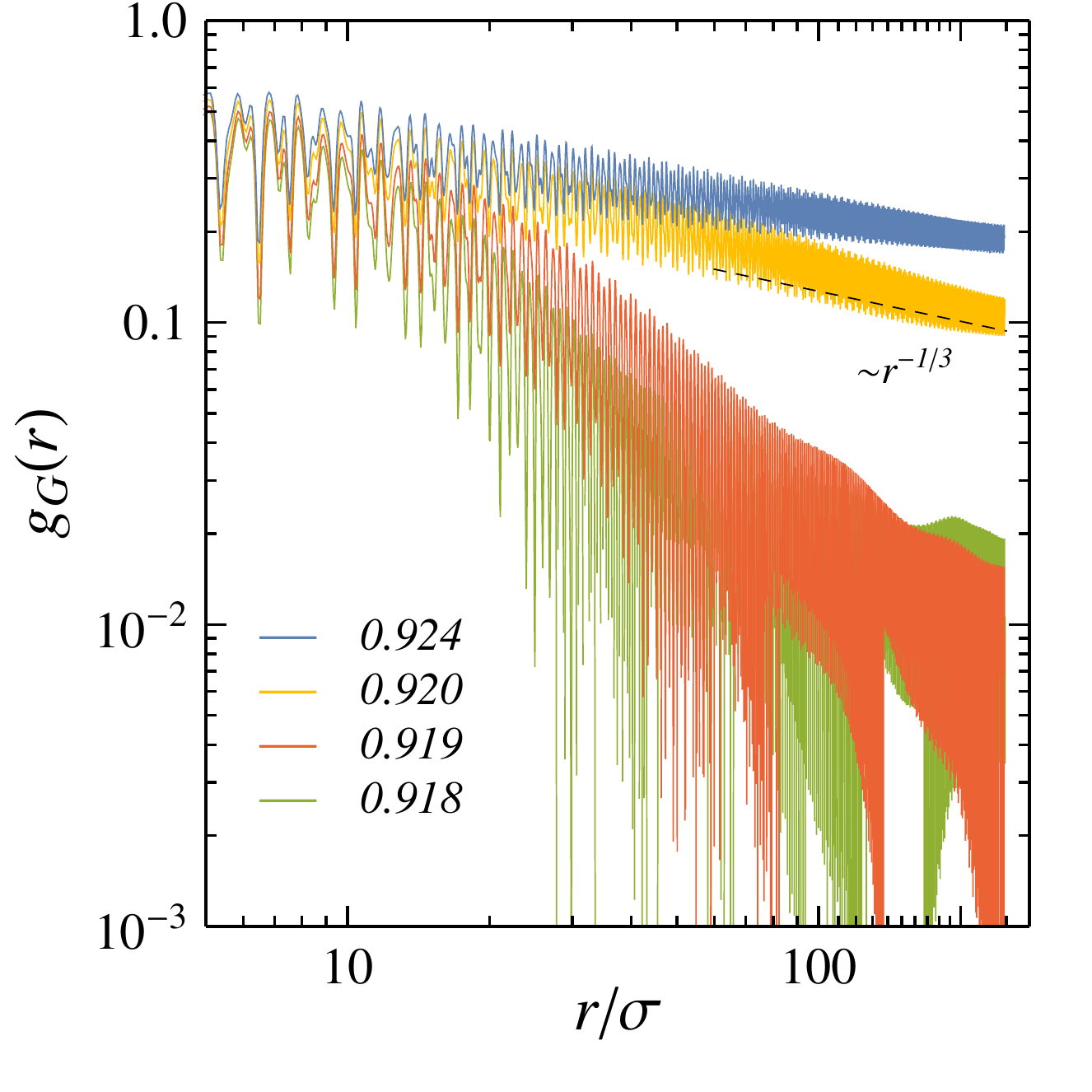}
    \caption{Positional correlation function $g_{G}(r)$ for
      a system of $N=512^2$ particles at densities $\r \s^2$ specified in
      the legend. It shows power law decay $r^{-\eta_G}$ with 
      $\eta_G< 1/3$ at densities $\r \s^2 > 0.920$. At the solid melting point, 
      $\r \s^2=0.920$, the correlation shows a power
      law decay $r^{-\eta_G}$ with an exponent $\eta_G=1/3$, indicated by the
      dashed line $r^{-1/3}$.
      At further lower densities, $\r \s^2 <0.920$, the
      correlation shows exponential decay. }
      \label{fig:psiG2corr}
\end{figure}

\begin{figure}[!ht]
  \centering
  \includegraphics[width=\linewidth]{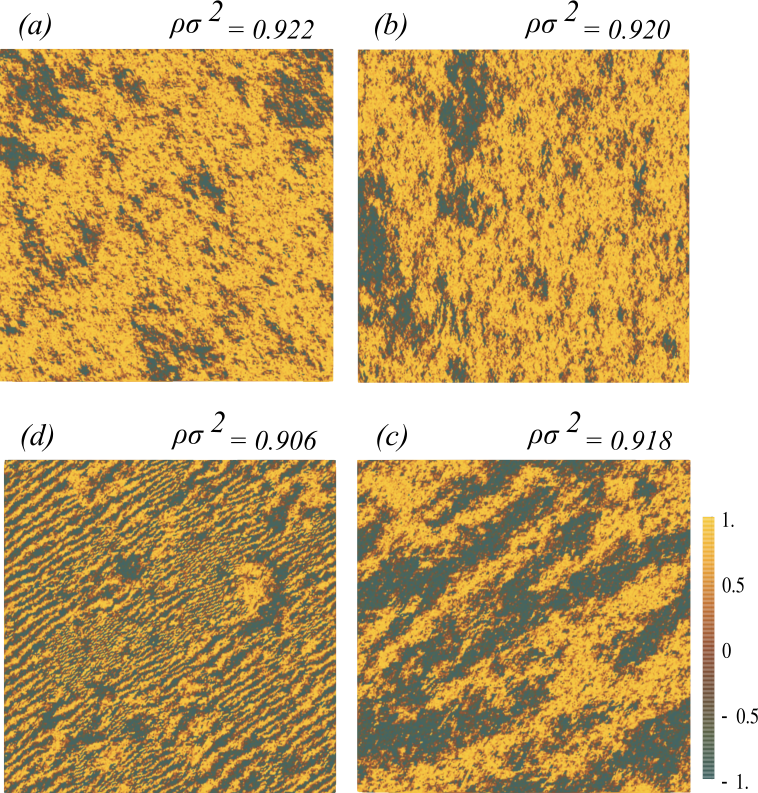}
  \caption{Plots of the positional order $\chi_i $
    for densities indicated above each plot. In the solid phase,
    $\r \s^2 \geq 0.920$, the plots show high positional order~(($a$) and ($b$)). 
    Immediately below the solid melting point, at $\r \s^2=0.918$,
    alternating broad bands of parallel and anti-parallel
    alignment of $\chi_i$ appear in the system~($c$). As the
    density is further reduced, these bands get narrower and shows up in the form of 
    alternating thin stripes.  }
  \label{fig:local_solid_order}
\end{figure}

\subsubsection{Finite size scaling}

The change in the nature of order across the solid melting can be examined using
a finite size scaling analysis. 
We calculate the solid order parameter $\la \psi_{\Gv} (\ell) \ra$ 
over sub- systems of varying size $\ell^2 = \ell_x \times \ell_y$. 
The solid order calculated
over the whole system is denoted here by $\la \psi_\Gv (L) \ra$. In Fig.~\ref{fig_psiG2scaling}
we show plots of 
$\la \psi_\Gv(\ell)\ra / \la \psi_\Gv(L)\ra$ 
with system size $\ell/L$, at different mean densities $\r \s^2$ across the solid melting. 
The solid order parameter corresponding to the QLRO solid phase is expected to show a power law decay with system size, $\la \psi_{\Gv}(\ell)\ra \sim \ell^{-\eta_G}$. Within the KTHNY theory, 
the exponent $\eta_G$ approaches $\eta_G^\ast = 1/3$ from below, as the solid approaches the melting
point. After melting, $\la \psi_{\Gv}(\ell)\ra$ is expected to show an exponential decay with $\ell$, characterizing the short ranged order.  
The decay in $\la \psi_{\Gv}(\ell)\ra$ with $\ell$, as shown in Fig.~\ref{fig_psiG2scaling}, is consistent with this picture. At densities higher than the melting point $\r \s^2=0.920$, $\la \psi_{\Gv}(\ell)\ra$ decreases with $\ell$ as a power law with exponent $\eta_G < 1/3$. At $\r \s^2 < 0.920$ the system- size dependence of $\la \psi_{\Gv}(\ell)\ra$ shows a stronger exponential decay.

\subsubsection{Positional correlation}
The change in order is further characterized by the positional correlation 
$g_{G}(r) = \la e^{i \Gv \cdot \rv_{ij}} \d(r - r_{ij}) \ra$, where $\rv_{ij} = \rv_i - \rv_j$
is the inter- particle separation vector, $r_{ij} = |\rv_{ij}|$, and $\Gv$ denotes the 
the reciprocal lattice vectors corresponding to the six quasi- Bragg peaks.
To explore the power law nature of the correlation in QLRO solid over a longer length scale, 
we use system size $N=512^2$ in plotting $g_{G}(r)$ in Fig.~\ref{fig:psiG2corr}. 
It is plotted at different densities across the melting transition. The change in the decay in correlation 
from power law $g_{G}(r) \sim r^{-\eta_G}$ in the QLRO solid phase at $\r \s^2 = 0.924$,
to exponential after melting $\r \s^2 \leq 0.919$ is clearly observed. At the melting point 
$\r\s^2 = 0.920$, the correlation shows power law decay consistent with the KTHNY 
prediction $r^{-1/3}$.

\subsubsection{Local positional order}
The local positional order can be visualized using 
$e^{i\,\Gv \cdot \bf{r}_i}$, where $\rv_i$ denotes the position
vector of $i$-th particle. This is a unit vector in two dimensions, and plotted for typical equilibrium configurations in Fig.~\ref{fig:local_solid_order} in terms of a heat map of the projection $\chi_i= e^{i\,\Gv \cdot \bf{r}_i} \cdot {\bf \hat x}$ with respect to the $x$-axis, ${\bf \hat x} = (1,0)$. 
In calculating this, we use the peak position of structure factor in the first quadrant. 
The fluctuations of $\chi_i$ is small in the solid phase. The positional order remains
mostly oriented over the system size~(Fig.~\ref{fig:local_solid_order}~($a$) and ($b$)). The
appearance of small patches with anti- parallel orientation of positional order is associated with the QLRO nature of the solid. After melting, the positional order gets short- ranged. 
This can be seen from the
formation of striped patterns at $\r \s^2=0.918$~(Fig.~\ref{fig:local_solid_order}~($c$)). The size of the stripes corresponds to the correlation length over which the order remains oriented. The widths of the stripes get narrower with the reduction of correlation length at lower density, as can be seen from the plot at $\r \s^2=0.906$ in Fig.~\ref{fig:local_solid_order}($d$). 

\begin{figure}[!t]
\centering
\includegraphics[width=7cm]{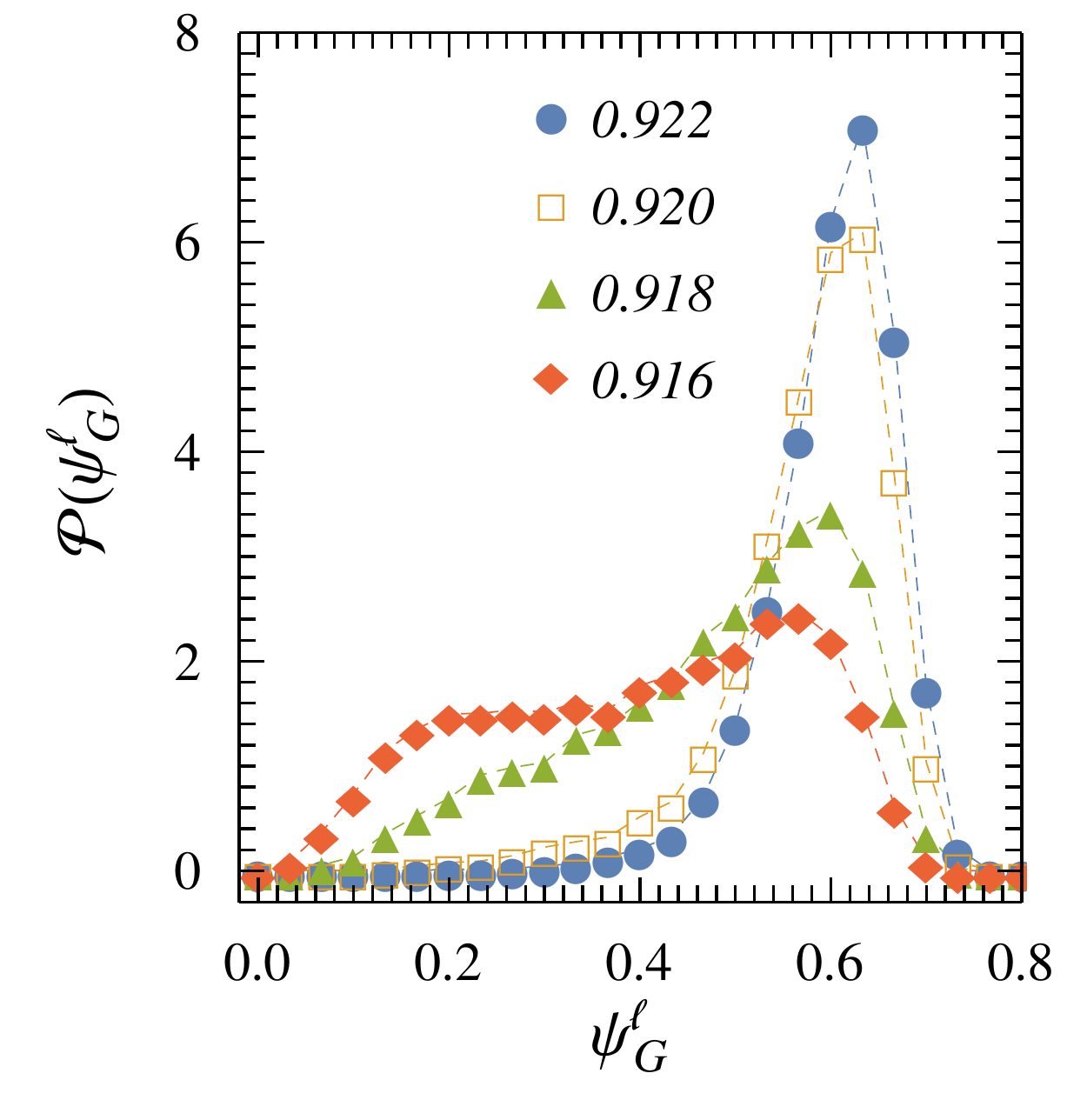}
\caption{Probability distribution of the solid order parameter ${\cal P}(\psi_\Gv^\ell)$ calculated
over sub-systems of size $\ell^2=L_x/10 \times L_y/10$, for different densities as
  indicated in the legend. The distribution remains unimodal across the solid melting point 
  at $\r_m \s^2=0.920$. At a lower density the peak of the distribution shifts to a lower $\psi_G^\ell$. 
}
  \label{fig:psiG_hist}
\end{figure}

\subsubsection{Distribution function of solid order}

Here we calculate the solid order parameter $\psi_\Gv^\ell$ over subsystems of 
size $\ell^2=L_x/10 \times L_y/10$, and obtain their distribution function ${\cal P}(\psi_\Gv^\ell)$~(Fig.~\ref{fig:psiG_hist}). The distribution function remains unimodal across 
the melting point $\r \s^2=0.920$. The peak of the distribution 
shifts to lower values as the solid melts. 
This unimodal nature of the distri- bution function signifies the absence of any metastable phase on the other side of the transition, a characteristic feature of continuous transitions.

The results obtained in this section shows that the melting of solid is a continuous transition,
consistent with the KTHNY melting scenario. The structure factor shows that the solid melts to 
a hexatic. The solid melting point is obtained at a density $\r \s^2=0.920$ clearly separated from, and larger than the coexistence interval obtained from the Mayer-Wood loop in the equation of state.   

\subsection{Hexatic melting}

\begin{figure}[t]
\centering
{\includegraphics[width=7cm]{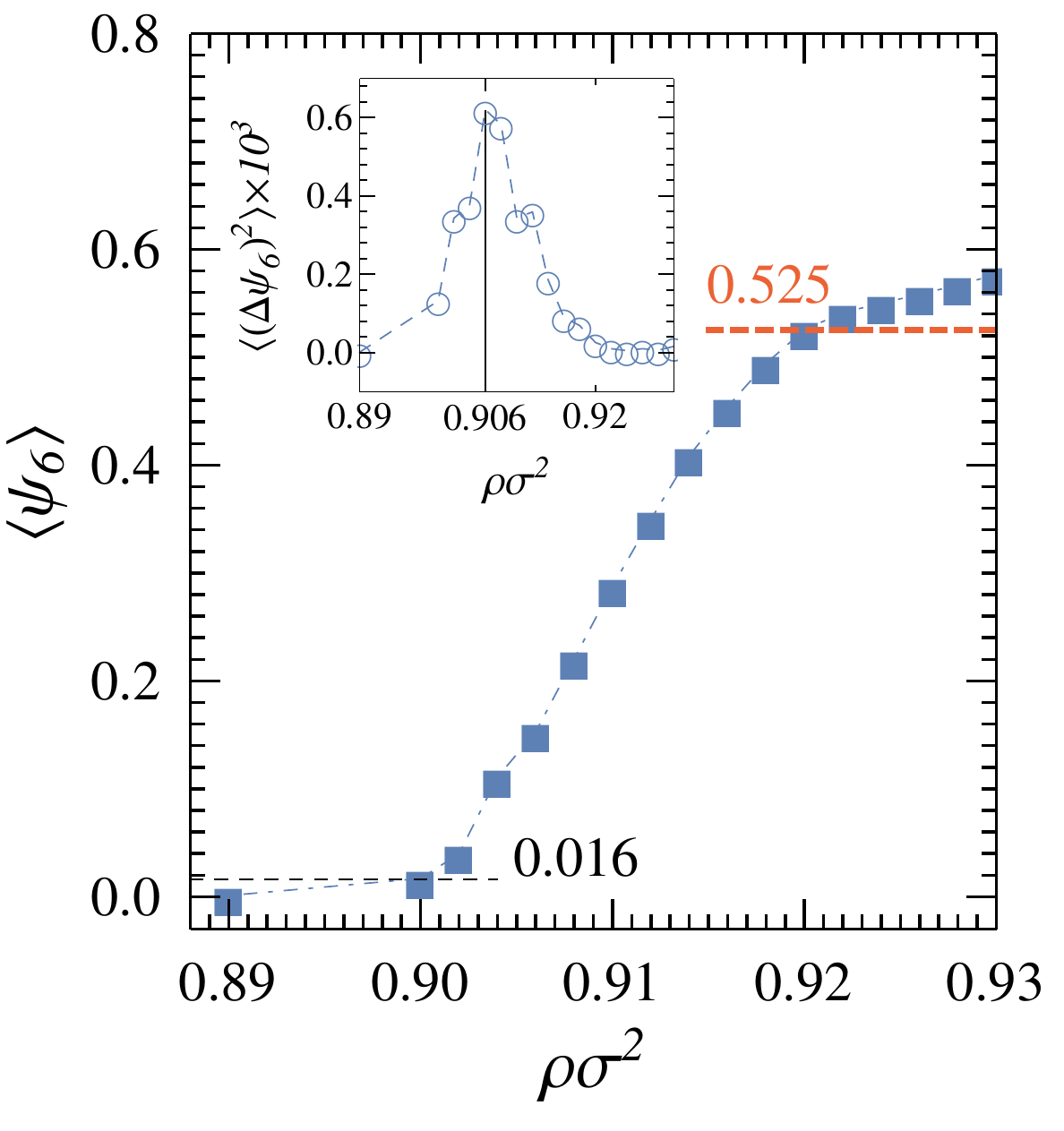}};
\caption{Plot of the bond orientational order parameter and its fluctuation (inset)
  as a function density. The peak of the order parameter fluctuation
  appears at a density $\r \s^2=0.906$. }
  \label{fig:psi6_vs_dens}
\end{figure}

In Fig.~\ref{fig:psi6_vs_dens}, we plot the mean amplitude of the hextic order parameter
$\la \psi_6 \ra = \la\, \mid (1/N) \sum_{i=1}^N \psi_6^i \mid^2\, \ra$ as a function of density. The hexatic order reduces continuously with lowering of mean density to vanish near $\r \s^2 = 0.900$. 
The fluctuations in hexatic order, $\la \D \psi_6^2\ra$, plotted in the inset of Fig.~\ref{fig:psi6_vs_dens} 
shows a pronounced maximum at $\r \s^2 = 0.906$, identifying the hexatic melting point.   
Note that this melting point $\r \s^2 = 0.906$ is right inside the coexistence interval obtained from 
the Mayer-Wood loop. Thus the hexatic melting is associated with phase coexistence.  

\subsubsection{Finite size scaling}

\begin{figure}[!ht]
  \centering
  \includegraphics[width=7cm]{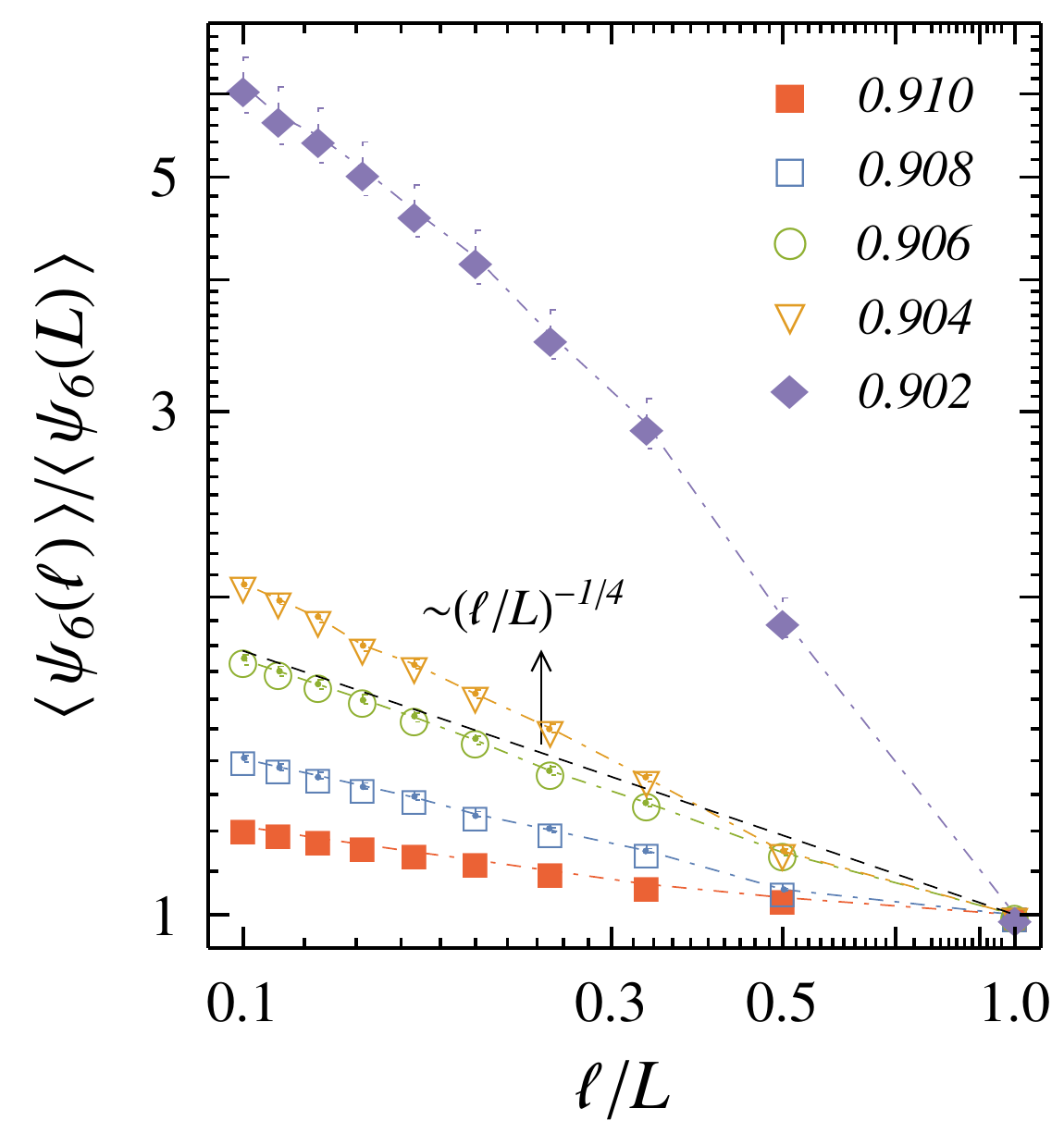}
   \caption{System size scaling of orientational order parameter
    $\psi_{6}$ for the densities indicated in the legend. The dashed
    line shows the power law $(\ell/L)^{-1/4}$.  }
  \label{fig:psi6scale}
\end{figure}

We use a finite size scaling analysis to establish the change in the nature of hexatic order across the hexatic- melting transition. We calculate the order parameter $\la \psi_6 (\ell) \ra$ 
over sub- systems of varying size $\ell^2 = \ell_x \times \ell_y$. The hexatic order calculated
over the whole system is denoted by $\la \psi_6 (L) \ra$. In Fig.~\ref{fig:psi6scale}
we show plots of $\la \psi_6(\ell)\ra/\la \psi_6(L)\ra$ with system size $\ell/L$, at different densities. 
Above the hexatic melting point, $\r \s^2 \geq 0.096$, the ratio shows power law decay $\la \psi_6(\ell)\ra/\la \psi_6(L)\ra \sim \ell^{-\eta_6}$. The single hexatic phase is characterized by such power laws. The KTHNY theory predicts that $\eta_6$ approaches $1/4$ from below as one approaches the hexatic melting point. Fig.~\ref{fig:psi6scale} shows apparent consistency with such expectation. At lower densities it transforms to exponential decay.  

However, the apparent power- law decay at the hexatic melting point is not consistent with the evidence of phase coexistence obtained from the Mayer- Wood loop, and the local heat- map of hexatic order parameters in Fig.~\ref{fig:qualitative_description}~($a$). We note that, in finite sized simulations~($N=256^2$ here),  if hexatic clusters span the system size even at coexistence, $\la \psi_6(\ell)\ra$ may display such an apparent power-law decay governed by the hexatic domains. In large enough systems, the domain boundaries between hexatic and fluid phases modfies the behavior leading to exponential tail. In the following, we show this in the hexatic correlation function, utilizing a larger system size.

\begin{figure}[!ht]
    \centering
    \includegraphics[width=7cm]{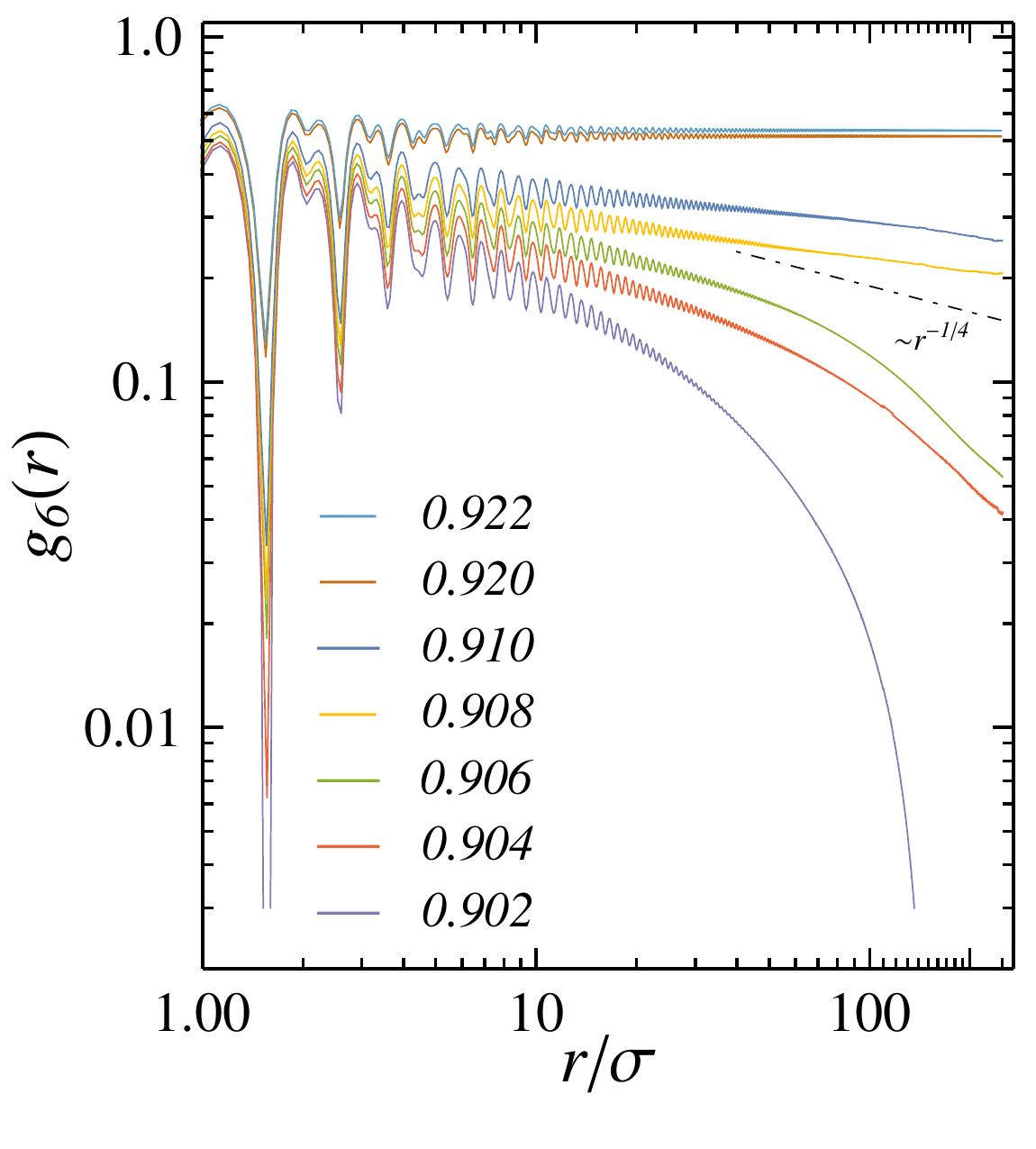}
    \caption{Plots of the hexatic correlation function $g_6(r)$
      for a system size $N=512^2$ at densities $\r \s^2$ indicated
      in the legend. The dot-dashed line is a plot of $r^{-1/4}$.
      }
    \label{fig:psi6corr}
\end{figure}  

\subsubsection{Hexatic correlation} 

In Fig.~\ref{fig:psi6corr} we plot the hexatic correlation $g_6(r) = \langle \psi^{i *}_{6} \psi^{j}_{6}\delta(r-r_{ij}))\rangle$,  where $\psi_6^{i *}$ denotes the complex conjugate of $\psi_6^i$, the
local bond orientational order and $r_{ij}$ is the separation between particles $i$ and $j$. 
In the solid phase $g_6(r)$ is expected to remain
independent of $r$, due to a long- ranged hexatic order. 
The QLRO single hexatic is expected to show a power law decay
$g_6(r) \sim r^{-\eta_6}$. Within the KTHNY theory $\eta_6$ is expected
to approach $1/4$ from below as one approaches the hexatic melting point,  
and after melting $g_6(r)$ is expected to show exponential decay. 

To investigate the power law nature of $g_6(r)$, and deviations from it, we use a larger system size of  $N=512^2$~(Fig.~\ref{fig:psi6corr}). 
At densities above the solid melting point $\r \s^2 = 0.920$, the correlation remains unchanged over the whole system size, a characteristic of the solid phase having long ranged hexatic order. 
At lower densities, but above the hexatic melting point, the correlations show power- law decay characteristic of hexatic phase. 
At the hexatic- melting transition $\r \s^2 = 0.906$, note that the correlation shows an apparent  power-law like decay $g_6(r) \sim r^{-1/4}$  up to $r \lesssim 70\s$, which crosses over to exponential decay for longer separations $r$.   
Thus simulations with system sizes smaller than the above- mentioned cross- over length, may show apparent power-law decay of $g_6(r)$ which could appear to be consistent with KTHNY theory. This we have checked separately for smaller systems, $N=128^2$ and $256^2$~(data not shown).  
However, the observed exponential tail in $g_6(r)$ of large systems is due to the presence of coexisting hexatic and fluid domains.
At densities lower than the hexatic melting point $\r \s^2 < 0.906$, the correlation shows exponential decay characterizing the fluid phase.

\subsubsection{Distribution function of hexatic order: phase coexistence}
\begin{figure}[!ht]
  \centering
  \includegraphics[width=7cm]{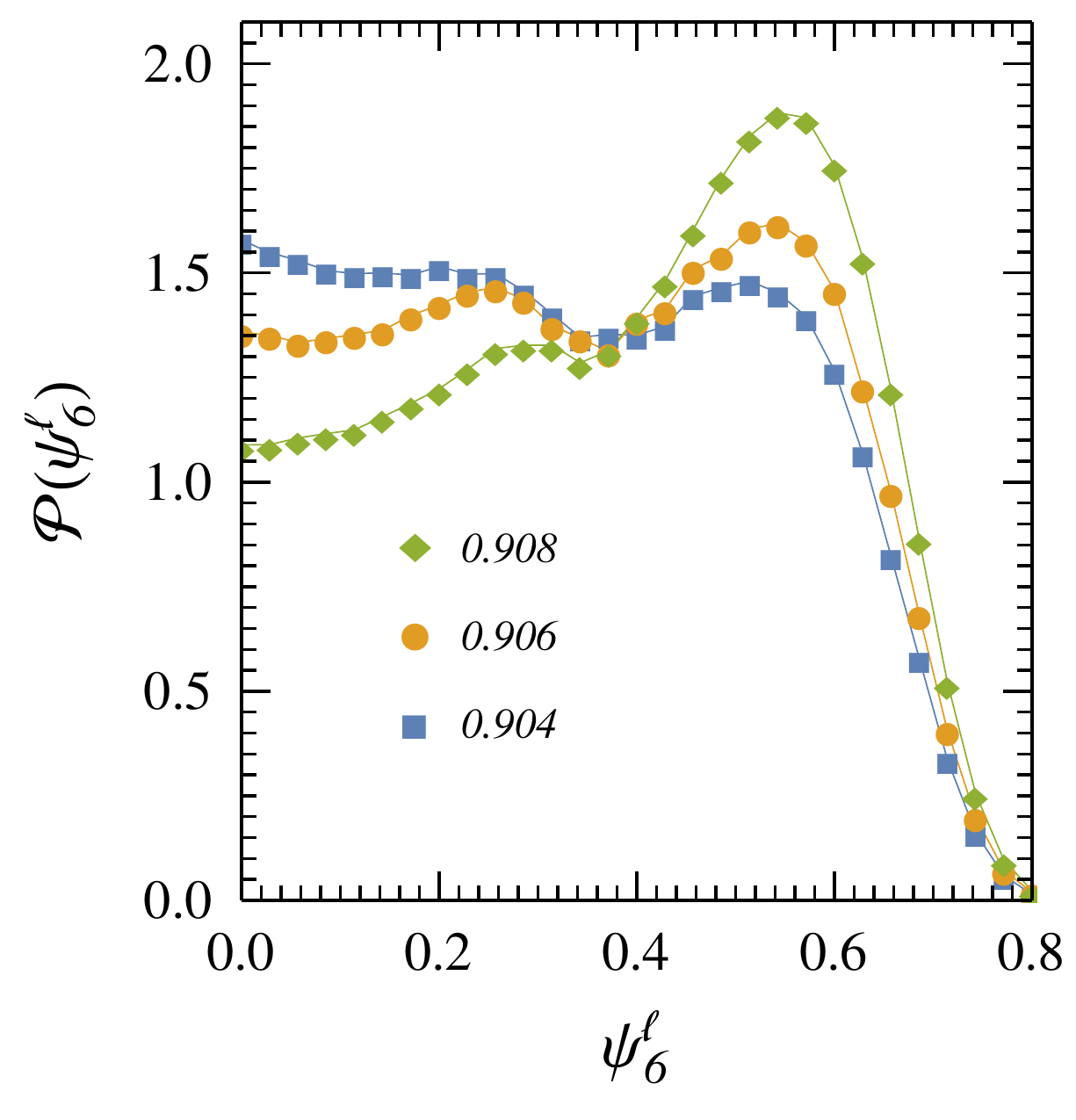}
  \caption{Probability distribution of the coarse- grained orientational order parameter
    ${\cal P}(\psi_6^\ell)$ at densities indicated in the legend.  }
\label{fig:block_psi6_hist}
\end{figure}

The nature of the hexatic melting is further characterized in terms of   
the probability distribution ${\cal P}(\psi_6^\ell)$ of local heaxtic order denoted here by 
$\psi_6^\ell := \psi_6(\ell)$, calculated over subsystems of size 
$\ell^2 = (L_x/10) \times (L_y/10)$. 
The distribution function at and around the hexatic- melting point $\r \s^2=0.906$
clearly show pronounced multi-modality due to phase coexistence~(Fig.~\ref{fig:block_psi6_hist}). 
The presence of metastable maximum in ${\cal P}(\psi_6^\ell)$ across the hexatic melting is a characteristic of the first order phase transition. 
At $\r \s^2=0.906$, the peak near $\psi_6^\ell = 0.56$ corresponds to the hexatic regions, while that near $\psi_6^\ell = 0.25$ is due to the coexisting fluid phase. 

Thus we have shown that the hexatic melting in WCA system is clearly a first order transition characterized by the coexistence of hexatic and fluid phases. The system remains in pure hexatic phase in the range of densities, $0.918<\r \s^2 < 0.920$,  lower than the solid- melting point, and above the upper limit of the coexistence interval. At lower densities, $0.900<\r \s^2 < 0.918$, the system gets into hexatic- fluid coexistence, with the hexatic melting point identified at $\r \s^2=0.906$. Below  $\r \s^2 = 0.900$ the system gets into the pure fluid phase.  

\begin{figure*}[!ht]
  \centering
  \includegraphics[width=\linewidth]{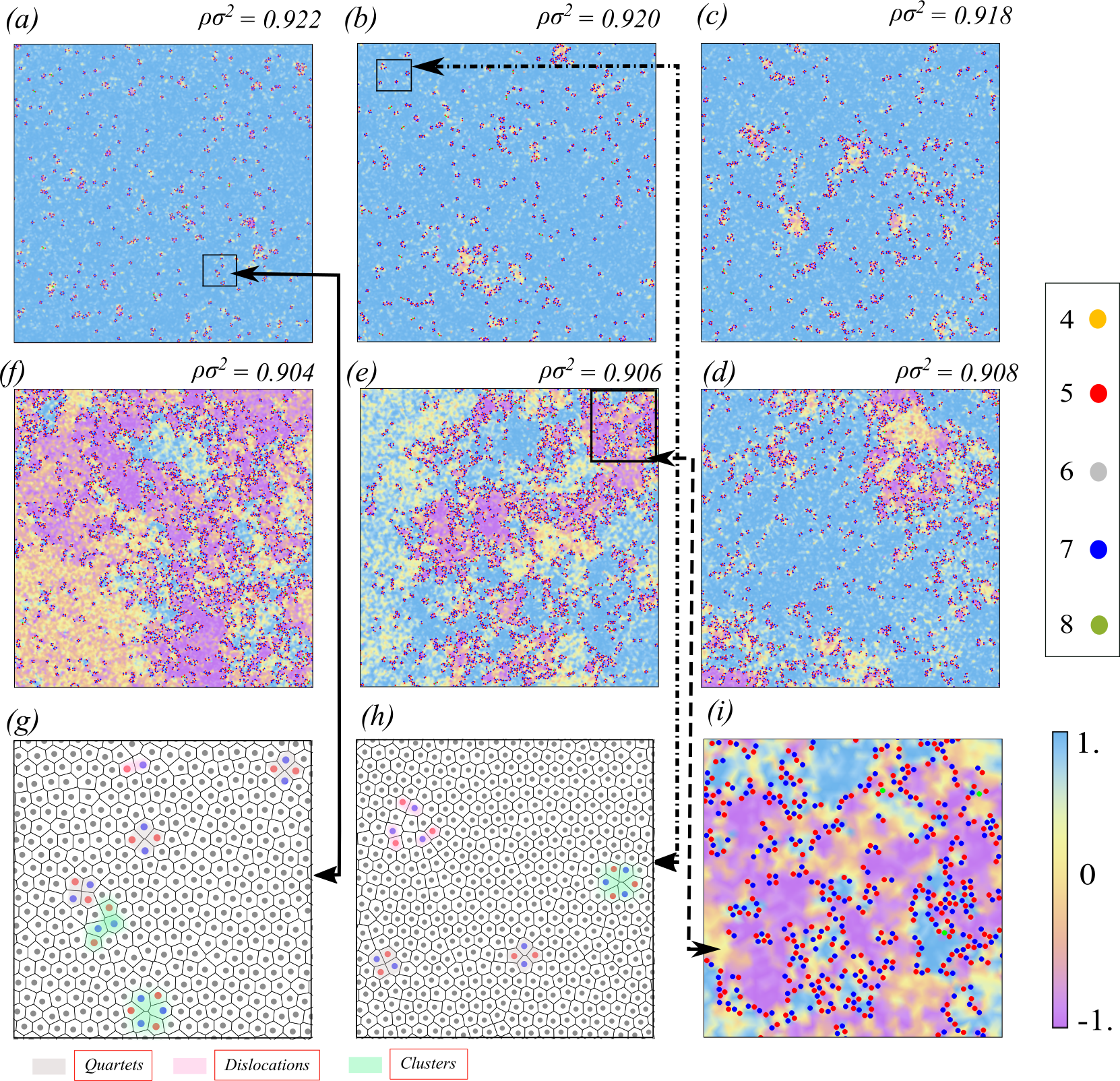} 
  \caption{{($a$)}--{($f$)}: Plots of the hexatic orientation $h_i = \psi_6^i \cdot {\bf \hat x}$
    in a region of size $200\s \times 200\s$. Superimposed on these plots, we show the
    particles with topological neighbors 
    $\nu=4$ (blue), $5$ (black), $7$ (green) and $8$ (red).
    {($g$)}--{($i$)}: Close up views of the regions
    indicated in the panels ($a$), ($c$) and ($e$) are shown to highlight the locations of $\nu$- fold 
    defects. The particles with $\nu=6$  topological neighbors are indicated by gray.  
    In ($g$) and ($h$), the Voronoi tessellation is shown, and the quartets, dislocations 
    and clusters of defect are highlighted by shading the corresponding Voronoi cells with gray, pink, and green, respectively. In ($i$), strings of defect are observed at 
    the interfaces between hexatic and disordered fluid.  }
\label{fig:local_solid_heatmap}
\end{figure*}

\subsection{Topological defects}
\label{sec:defects}

In this section, we discuss defect formation as the system undergoes the two stage melting from solid to hexatic to fluid phase. 
We use Voronoi tessellation to identify the topological neighbors of a given particle. 
In a perfect triangular lattice, each particle has $\nu=6$ neighbors. A particle with 
$\nu \neq 6$ neighbors is identified as a $\nu$-fold defect, e.g., $\nu=5$- or $7$- fold defects.  
Fluctuations in solid can accommodate $5-7-5-7$ bound quartets, corresponding to bound dislocation- anti-dislocation pairs.
In the previous section, we established a continuous melting of solid to hexatic. 
Within the KTHNY theory, solid- hexatic melting transition is mediated by     
unbinding of $5-7$ and $7-5$ pairs signifying unbinding of dislocations, line defects in 2d solid. 
Within the same theory, hexatic melting is expected to be continuous and mediated by unbinding of dislocations to free $5$-fold or $7$-fold disclinations. In contrast, as we have shown the hexatic- fluid melting for WCA system is a first order transition characterized by phase coexistence.   

In Fig.~\ref{fig:local_solid_heatmap} we show defect
formation along the melting
transitions. Each figure in Fig.~\ref{fig:local_solid_heatmap}($a$)-($f$) shows a heat map of the 
hexatic order for each particle projected along the $x$-axis, 
$h_i := \psi_6^i \cdot {\bf \hat x}$, where the vector ${\bf \hat x} =  (1,0)$.
By definition, for a perfect triangular lattice 
$\psi_6^i = (\Re \psi^i_6,\operatorname{Im} \psi_6^i) = (1,0)$.
Thus the blue regions in the heat maps denote large hexatic order $h_i \approx 1$. 
The system shows a high degree of hexatic alignment for $\r \s^2 \geq 0.918$ as is shown 
in Fig.~\ref{fig:local_solid_heatmap}($a$)-($c$). 
The patches of other colors signifying disordered fluid droplets, grow to significant fraction of the 
system size at lower densities, see Fig.~\ref{fig:local_solid_heatmap}($d$)-($f$).  
The interfaces between coexisting domains can be identified by noticing the change in color.    
In addition, we have plotted the $\nu=4,\,5,\,7,\,8$- fold defects on the heat maps of $h_i$ in  Fig.~\ref{fig:local_solid_heatmap}($a$)-($f$). These are shown more clearly by focussing on small 
regions of the system in Fig.~\ref{fig:local_solid_heatmap}($g$)-($i$). 

In Fig.~\ref{fig:local_solid_heatmap}($g$),($h$), we show the Voronoi tessellation corresponding to densities $\r \s^2=0.922$, $0.920$, respectively. We identify the particles with five neighbors in red, and those with seven neighbors in blue. Moreover, we use shades of gray, green and pink to identify Voronoi cells corresponding to bound quartets, clusters, and dislocations (separated $5$-$7$ defect pairs). As can be seen from the plots, spatial positioning of the defects are related to locally low hexatic order $h_i$. 
Fig.~\ref{fig:local_solid_heatmap}($i$) magnifies a small region of  Fig.~\ref{fig:local_solid_heatmap}($e$) corresponding to the hexatic melting point $\r \s^2 = 0.906$. Fig.~\ref{fig:local_solid_heatmap}($i$) shows strings of defects located predominantly on the interfaces.  

\begin{figure}[htp]
\centering
\includegraphics[width=\linewidth]{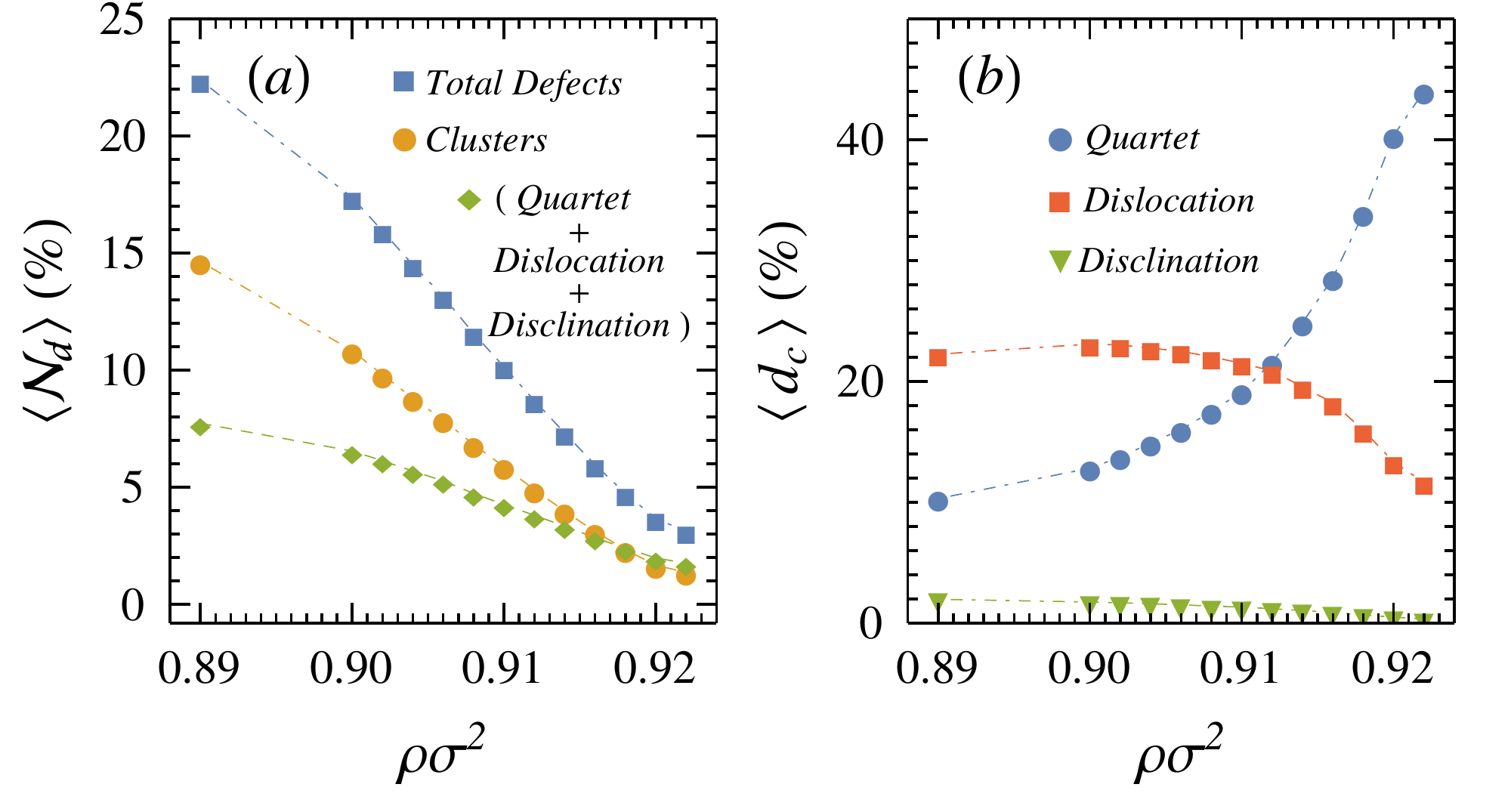}
\caption{ ($a$)~Variation of percentage defect-fractions $\la \mathcal{N}_d \ra$~($\Box$),
$\la \mathcal{N}_c \ra$~($\circ$), $\la \mathcal{N}_s \ra$~($\diamond$),   with
  system density $\rho\sigma^{2}$. 
  ($b$)~The percentage defect fraction $\la d_c\ra$ of quartets, dislocations and disclinations are shown as a function
  of the system density.  }
  \label{fig:defect_types}
\end{figure}

A quantitative analysis of defect formation in the system is obtained
from the estimated number of different types of defects by keeping
track of particles with $\nu \neq 6$ neighbors. In Fig.~\ref{fig:defect_types}($a$)
we show the mean percentage fraction of defects
$\la \mathcal{N}_d \ra=(1-\la n_6\ra/N) \times 100$ as a function of
density, where at each density the averaging is done over $200$ independent
configurations. 
This gives the estimate of the percentage fraction of all the defects.
In addition, we separately consider $\la \mathcal{N}_s \ra$, a similar percentage fraction of a total sum of 
quartets, dislocations and disclinations taken together. Further, we separately consider 
$\la \mathcal{N}_c \ra$, the percentage fraction of clusters and strings of defects. The quartets are not counted in clusters. Variation of all theses defect fractions, $\la \mathcal{N}_d \ra$, $\la \mathcal{N}_s \ra$ and  $\la \mathcal{N}_c \ra$ with density are shown in Fig.~\ref{fig:defect_types}($a$).
We clearly observe that above the melting point of the solid, $\r \s^2=0.920$, the percentage
fraction of all these defects are negligibly small. They increase with reduction of density,
and in the fluid phase, $\la \mathcal{N}_d\ra$ increases by almost an order of magnitude. 
The plot shows $\la \mathcal{N}_c \ra > \la \mathcal{N}_s \ra$ that clusters and string of defects 
dominate all through. 

Finally, to get a better insight into the relative role of specific kind of defects in the phase
transitions, we examine what is the contribution of quartets, dislocations and disclinations
in the total defect fraction $\la \mathcal{N}_d \ra$. We estimate this using the
quantity $\la d_c \ra=\la n_d/N \mathcal{N}_d \ra \times 100$, where $n_d$ denotes the number
of defects contributing to quartets,  dislocations 
and disclinations, respectively. In Fig.~\ref{fig:defect_types}($b$) we show their variation
with density. At high density solid phase the fraction of quartets dominate. This fraction decrease
monotonically with decreasing density. In contrast, the dislocation fraction remains low
in the solid, and increases with decreasing density to saturate to $\sim 20\%$.  The increase (decrease) in dislocation fraction (quartet fraction) across the solid melting point $\r \s^2 = 0.920$ is consistent with the KTHNY  picture of dislocation mediated melting, in which bound dislocation- anti-dislocation pairs (quartets) are expected to unbind forming free dislocations. On the other hand, unlike the KTHNY melting of hexatic,  disclinations,  the fraction of which remains low all through, play no significant role in the {\it first order} hexatic melting of the 2d WCA system.  
Rather, the monotonic increase and dominance in the fraction of defect- clusters across the hexatic melting~(see Fig.~\ref{fig:defect_types}($a$)) is due to the appearance of strings of defects at the interfaces of coexisting hexatic and fluid domains~(Fig.~\ref{fig:local_solid_heatmap}($i$)), a characteristic of the first order hexatic- fluid transition.

\section{Conclusion}
\label{sec_discussion}

In summary, we have presented a detailed study of the melting transitions of a system of particles interacting via the Weeks- Chandler- Anderson (WCA)  potential, using large scale molecular dynamics simulations. With reduction of density, the system shows two- step melting. The solid melts into a hexatic phase via a continuous transition at the density $\r \s^2 = 0.920$. This melting is associated with unbinding of dislocations, like the solid- melting within KTHNY theory. The system remains in the single hexatic phase only up to $\r \s^2 = 0.918$. The hexatic melts into a fluid at $\r \s^2 = 0.906$ via a first order phase transition, characterized by hexatic- fluid coexistence, which is unlike the KTHNY prediction. This melting is associated with formation of large strings and clusters of defects located mainly at the boundaries of coexisting domains. A wide range of densities $0.900 \leq \r \s^2 \leq 0.918$ shows this coexistence. Finally at densities $\r \s^2 < 0.900$ the system gets into a single fluid phase.

The solid melting is characterized using the structure factor, solid order parameter, positional order and correlation function. The probability distribution of the local solid order remains unimodal across the transition,  showing consistence with the continuous melting.  Finite size scaling of order parameter, and the behavior of the correlation function $g_G(r)$ shows consistency with the KTHNY prediction, particularly a power law decay $g_G(r) \sim r^{-1/3}$ at the solid- melting point. The hexatic melting is studied using the bond- orientational hexatic order parameter. The probability distribution of local hexatic order shows bimodality across transition, due to the presence of metastable state across the first order melting.  The correlation function of hexatic order $g_6(r)$ shows a power- law decay with $r$ within the hexatic phase, capturing its quasi- long- ranged nature of order. However, at the hexatic melting point $g_6(r)$ shows exponential tail due to the coexistence of hexatic and fluid, a behavior unlike the KTHNY prediction. The Mayer- Wood loop in the equation of state is due to the phase- coexistence at first order transition, and identifies the coexistence interval. The solid melting point remains clearly at a density higher than this interval. 

Thus we established a continuous solid- hexatic melting followed by a first order hexatic- fluid melting in the WCA system. The detailed analysis of different defect types and visualization of their locations showed that the continuous solid- melting is associated with dislocation unbinding, whereas strings of defects that localize near the hexatic- fluid domain boundaries dominate the first order hexatic melting transition.


\section*{Acknowledgements}
D.C. thanks ICTS-TIFR, Bangalore, for an associateship, and SERB, India for financial support through grant numbers MTR/2019/000750 and EMR/2016/001454. This research was supported in part by ICTS during a visit for participating in the program - 7th Indian Statistical Physics Community Meeting (Code: ICTS/ispcm2020/02).

\bibliographystyle{prsty} 

\end{document}